\documentclass[10pt]{article}

\usepackage{cite} 

\usepackage{graphicx}
\usepackage[font=scriptsize]{caption}

\begin{document}

\title{\textbf{Students' understanding of gravity using the rubber sheet analogy: an Italian experience}}
\date{}
\maketitle

\begin{center}
\author{A Postiglione$^{1,2}$, I De Angelis$^{1,2}$}\\
\end{center}

\paragraph{} \parbox[t]{1\columnwidth}{$^1$Dipartimento di Matematica e Fisica, Universit\`a degli Studi Roma \\Tre, Rome (ITALY)\\%
    $^2$INFN Sezione di Roma Tre, Rome, (ITALY)\\
    
    adriana.postiglione@uniroma3.it}

\begin{abstract}
General Relativity (GR) represents the most recent theory of gravity, on which all modern astrophysics is based, including some of the most astonishing results of physics research. Nevertheless, its study is limited to university courses, while being ignored at high school level. To introduce GR in high school one of the approaches that can be used is the so-called rubber sheet analogy, i.e. comparing the space-time to a rubber sheet which deforms under a weight. In this paper we analyze the efficacy of an activity for high school students held at the Department of Mathematics and Physics of Roma Tre University that adopts the rubber sheet analogy to address several topics related to gravity. We present the results of the questionnaires we administered to investigate the understanding of the topics treated to over 150 Italian high school students who participated in this activity.
\end{abstract}

\noindent{\it Keywords\/}: gravity, Einstein, General Relativity, space-time, Secondary Education, hands-on activity, experimental activity

\section{Introduction}
The most recent and successful theory describing gravity is the Theory of General Relativity (GR), introduced by Albert Einstein in 1916 \cite{Einstein}. This theory relies on the concept of space-time, a four-dimensional entity that unifies space (which has three dimensions) and time (one dimensional). According to this theory, the space-time change according to the objects placed in it: a massive object (a galaxy, a star, a planet, a dog) deforms the space-time producing gravity, i.e. attracting nearby masses.

All modern astrophysics, including some of the most important and recent discoveries of the field \cite{GravitationalWaves,GravitationalWaves2, GravitationalWaves3, ImageBH} is based on GR. But yet, its study is typically addressed only in advanced University courses, while it is ignored at lower levels of education, such as in high schools, where gravity is only described using the Newtonian theory.  

The reasons for this are many. The mathematical complexity of GR forces to adopt a purely qualitative approach, which is not easy to realize without oversimplifying the concepts too much. Moreover, even if one decides to only deal with the basic concepts at a qualitative level, such as the space-time and its deformation, he/she will have to accept a new vision of the world, far from the everyday experience.

In recent years, several efforts have been made in order to address these issues and create activities suitable to introduce GR in school curricula \cite{Possel,Zahn,Stannard,Velentzas, Kaur,Farr,Boyle,MagdalenaFreeFall,MagdalenaGravityWarpedTime,Hartle,Weiskopf}. Several other works focus on quantitative analysis of students' understanding of gravity and GR, ranging from very young kids to university students \cite{Pitts, Baldy, MagdalenaEvaluation, MagdalenaFourDimensions, Choudhary, Watkins, Graham, Williamson, Watts, Palmer, Gonen, Gousopoulos, Bandyopadhyay, Bandyopadhyay2}.
This present paper aims at contributing to this discussion through the analysis of  questionnaires administered to over 150 Italian high school students who participated to an activity we built to treat gravity and GR.

The model we used is the popular rubber sheet analogy (RSA), which compares the space-time to a rubber sheet that deforms under the weight of a mass, and allows to simulate the gravitational attraction through marbles and balls thrown on the warped sheet. Although this model shows some critical points extensively addressed in the literature \cite{MagdalenaUnderstandingCurved,Price,MagdalenaFreeFall,Gould}, it represents a powerful activity \cite{Possel, Farr, Baldy, Thorne} that has already demonstrated to be very well welcomed by Italian high school teachers \cite{our_paper}.

For this reason, at the Department of Mathematics and Physics of Roma Tre University we built a structure that could support a lycra sheet, and we used it as RSA to realize an activity addressed to high school students that could deal with different topics such as Kepler’s laws, gravity assist, gravitational lensing and black holes. We asked all participants to answer three questionnaires, one before the activity, one immediately after and one four months later. In this way we investigated their understanding of the most important aspects addressed during the activity, and the possible presence and persistence of misconceptions or wrong beliefs.  

The remaining paper is organized as follows. In section 2 we briefly describe the structure we used to exploit the RSA, the activity and the corresponding questionnaires. In section 3 we illustrate the results of our research, focusing on the main aspects that come out. In section 4, we discuss our achievements and in section 5 we present our conclusion and suggest some future development of our work. 

\section{Background and data collection}
\begin{figure}[ht]
  \centering
\includegraphics[width=0.5\columnwidth]{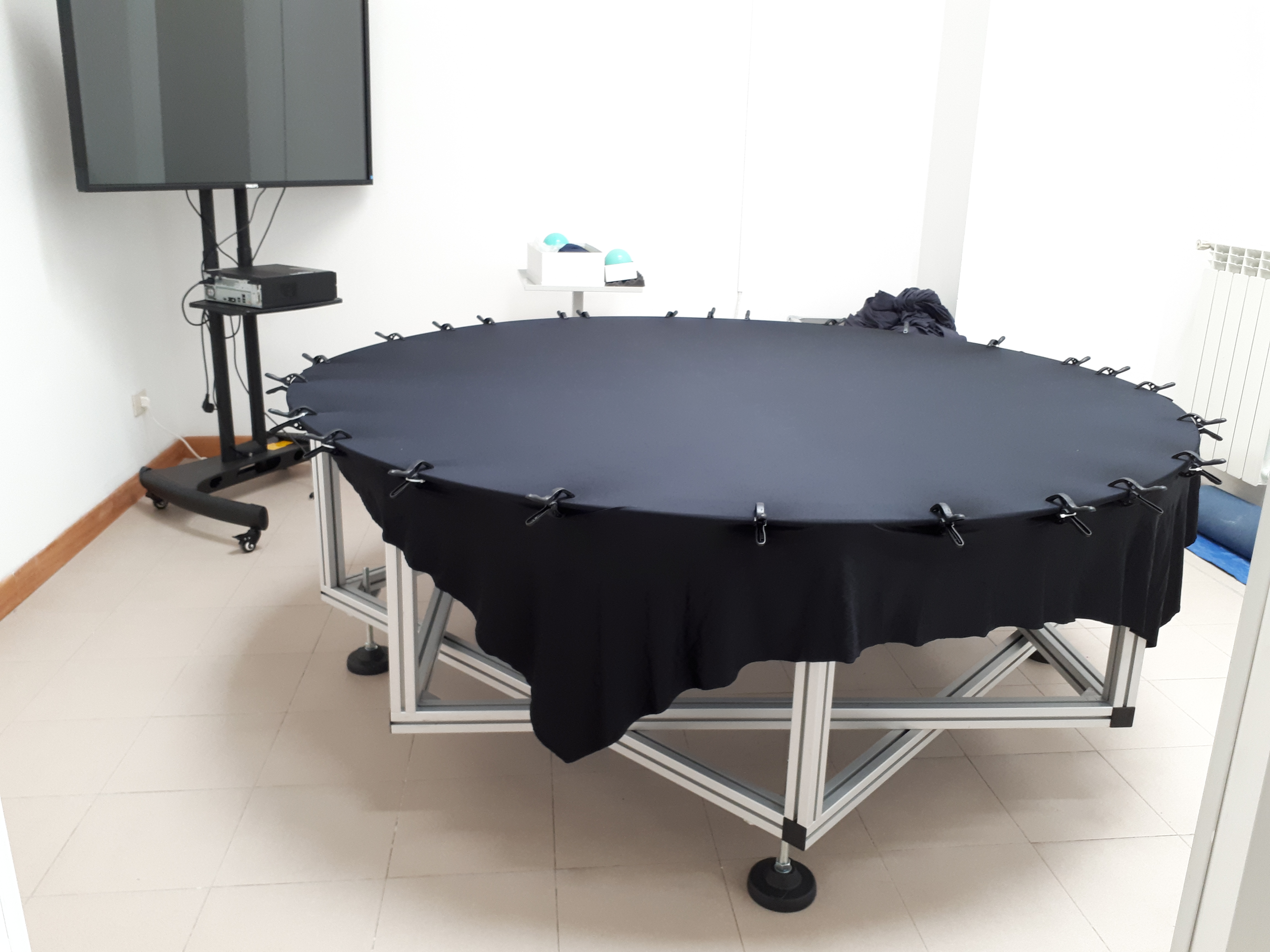}
\caption{The 1.8 m diameter structure of space-time we built at the Department of Mathematics and Physics of Roma Tre University in collaboration with the mechanical shop of the INFN Roma Tre Section.}
\label{fig:space-time}
\end{figure}

In order to carry out an activity that dealt with gravity using the RSA we built a circular structure of 1.8 meters in diameter in aluminum covered by a lycra sheet of about 2x1.5 meters (\textbf{Figure \ref{fig:space-time}}), in collaboration with the mechanical shop of the INFN Roma Tre Section. We chose this size for the structure to ensure that a group of about 25 people (a typical Italian high school class) could comfortably watch what showed on the sheet.

In the period January - February 2020 we used this structure to carry out our activity with 6 high school classes, engaging more than 150 students. 
Before, after and four months after the activity, we asked the participants to answer three questionnaires in order to explore their understanding of the topics addressed\footnote{The answers to our questionnaires were anonymous. This study was carried out in accordance with the principles outlined in IOP Science ethical policy.}.

\subsection{Description of the activity}
The activity consists of a lesson lasting an hour and a half during which moments of strong interaction with participants using the rubber sheet are alternated to theoretical insights using videos and photos.
We start with a description of the model of RSA: we first introduce the concept of space-time and how it is related to gravity, using a central weight and some marbles; we then focus on the simplifications that the usage of the rubber sheet implies, that can lead to misconceptions and wrong beliefs. We believe that bringing these misconceptions to light helps participants to overcome them (an idea that has been confirmed by the results of the questionnaires, as we will see in the following). First of all, we underline that space-time can be deformed by any mass (or energy) and not only by big masses. Then, we clarify that the rubber sheet represents a two-dimensional space-time. We thus point out that the space-time curvature originated by a spherical object is symmetrical in all directions, and therefore that there is no privileged direction in the Universe, contrary to what both the rubber sheet and the daily experience on Earth could lead to think. In other words, there is not an \textit{up} and \textit{down}, but only a \textit{near} or \textit{far}  from the source of gravity. 

After these first clarifications, we start to show the participants how this model, although simplified, can allow to visualize quite faithfully the way in which planets orbit the Sun, i.e. following Kepler's laws. Once the basic rules of the game are shown, we treat other examples of motion of celestial objects. We show the Earth-Moon system, the orbits of a planet around two stars and the phenomenon of gravity assist, that can explain the typical voyage of a space probe. It is worth reminding that throughout these activities the students are actively involved, and firsthand experience the behaviour of the marbles throwing them on the sheet: in this way their attention is kept high, and they become more willing to ask questions and join the discussion.

At this point a fairly complete picture of how the space-time and its deformation describe the phenomena that the students have only studied in terms of the Newtonian gravity has been given. Then we focus on the topics fully explained only by GR. We start from the phenomenon of gravitational lensing, representing light with a marble that deflects its trajectory when approaching the central weight. Through videos and photos, we then show the participants the consequences of this phenomenon, and how it is used by astrophysicists to characterize some celestial objects. 
We then introduce black holes, explain that they are compact objects, underline the fact that they strongly affect only the surrounding region and restate that this attraction does not point \textit{downwards}. Finally we talk about the gravitational waves, the way in which scientists have discovered them and their usefulness in improving our knowledge of the Universe.

\subsection{Description of the questionnaires}
We use  three questionnaires: one administered before the activity in order to analyze the prior knowledge of the participants, one administered right after the activity so that we could investigate the knowledge acquired thanks to the activity, and one questionnaire administered four months after the activity in order to study the permanence of the knowledge obtained.

All the questionnaires have the same structure: a first part that deals with the age and the school attended by the participants, a second part that focuses on the Newtonian theory of gravity and a third part that considers more complex topics related to GR.

When designing the answers to the questionnaires we paid attention in adding distractors so that we could investigate the presence of wrong beliefs and misconceptions. 

While the two questionnaires administered after the activity are identical, between the questionnaires administered before and after there is a slight difference: the pre-questionnaire includes both multiple choice questions with 5 alternatives (one of which was always \lq\lq I don't know\rq \rq) and open questions, while the post-questionnaires only used multiple choice questions with 5 alternatives, one of which could eventually be an open answer. 

\section{Results}
Overall we obtained 153 answers to the questionnaire administered just before the activity, 125 answers to the one administered just after the activity and only 42 to the one administered four months later the activity. This relevant decrease in the sample is mainly due to the fact that, unlike the other two, the last questionnaire was administered remotely, and in the period when the Italian schools were closed due to the first Covid-19 lock-down; the teachers could thus not follow their students' compiling live. 

\begin{figure}[ht]
  \centering
\includegraphics[width=0.85\columnwidth]{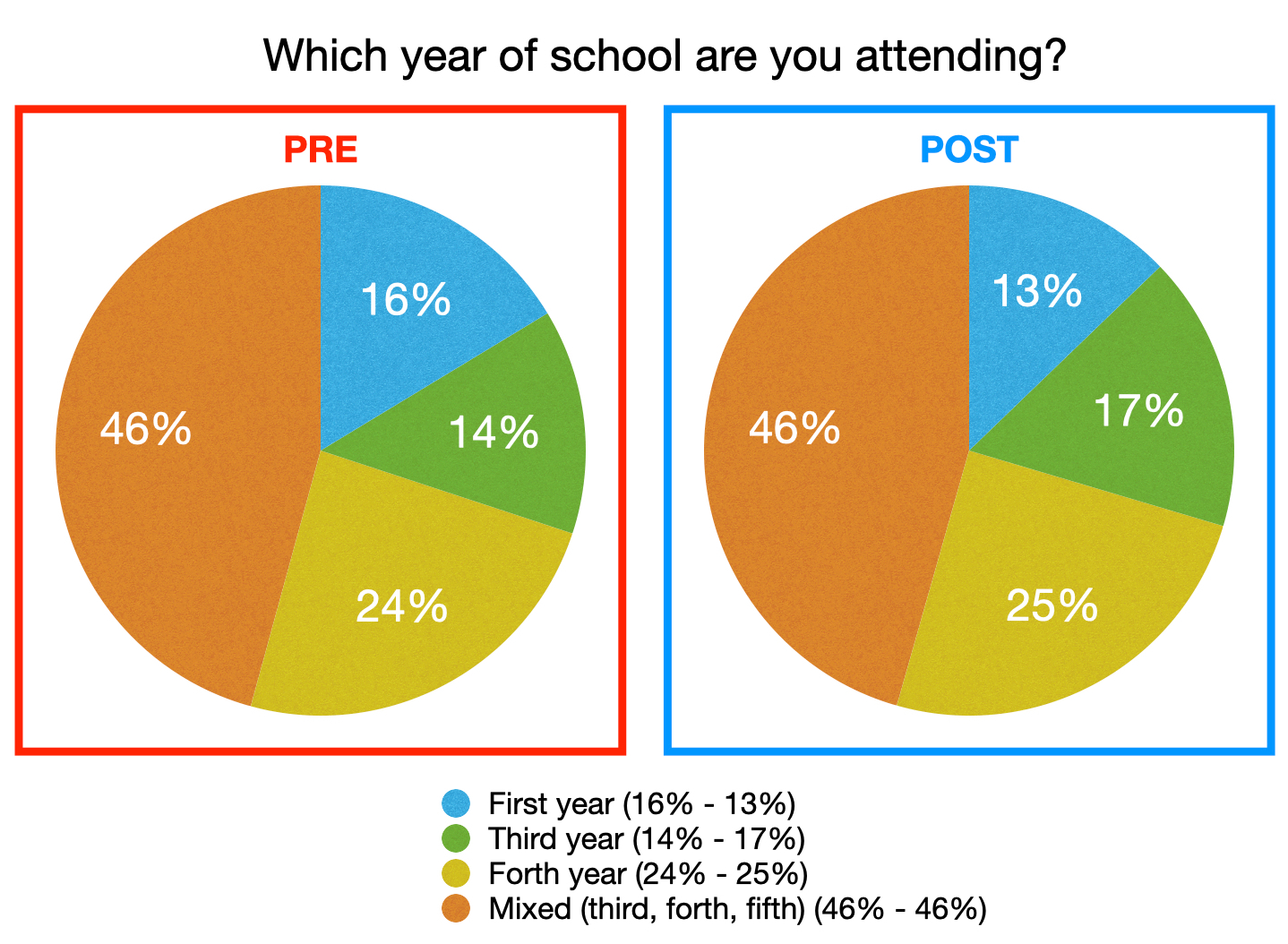}
\caption{\footnotesize{The age distribution of the participants before, after and four months after the activity. Since the Newtonian gravity is typically introduced during the third year, the majority of the participants already treated it. The percentage of each answer is shown in brackets (left: pre-questionnaire, right: post-questionnaire).}}
\label{fig:q0}
\end{figure}

\begin{figure}[ht]
  \centering
\includegraphics[width=0.85\columnwidth]{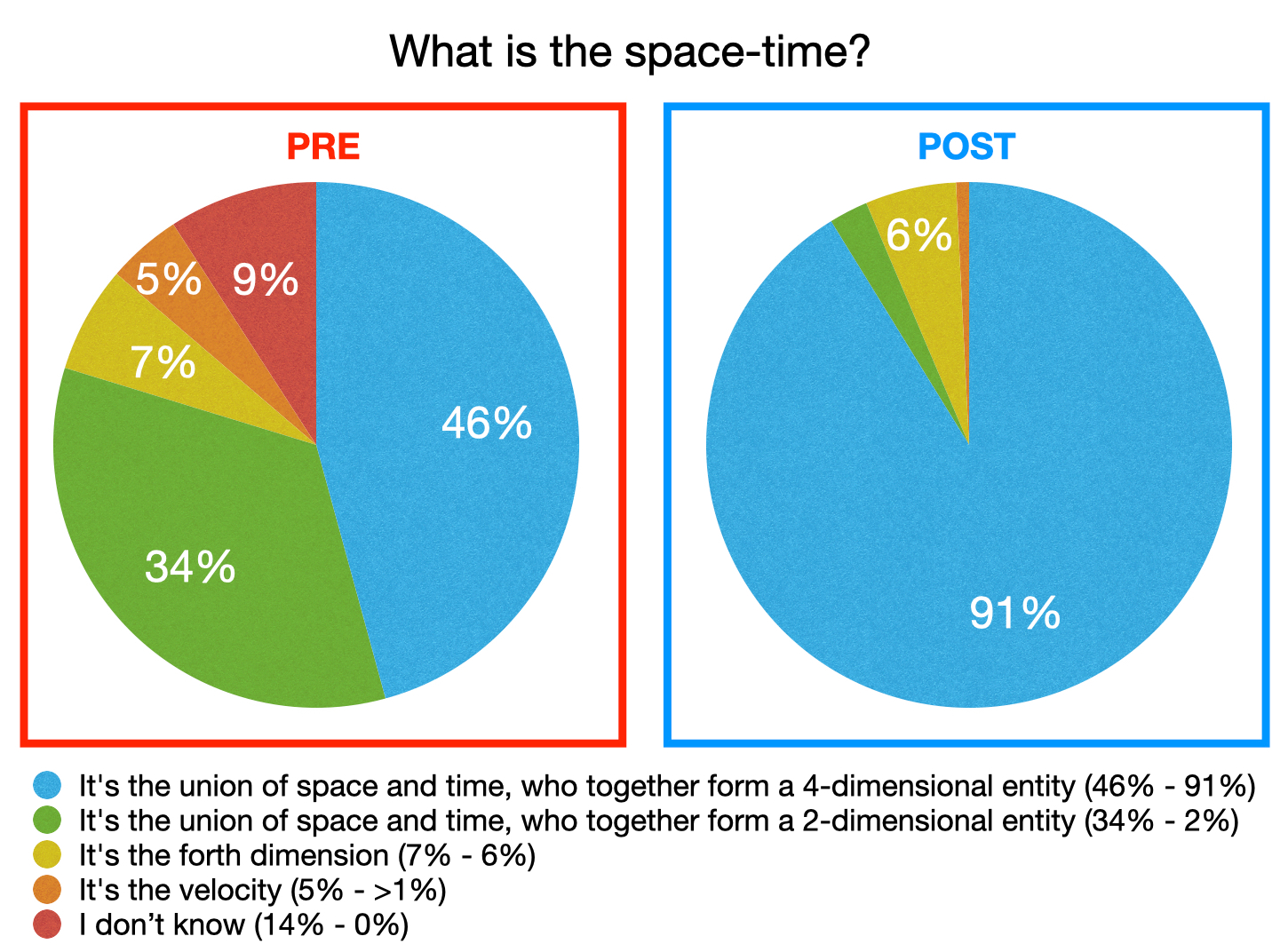}
\caption{The distribution of the answers to the question \textit{What is the space-time?}. The percentage of each answer is shown in the legend in brackets (left: pre-questionnaire, right: post-questionnaire).}
\label{fig:q1}
\end{figure}

\begin{figure}[ht]
  \centering
\includegraphics[width=0.93\columnwidth]{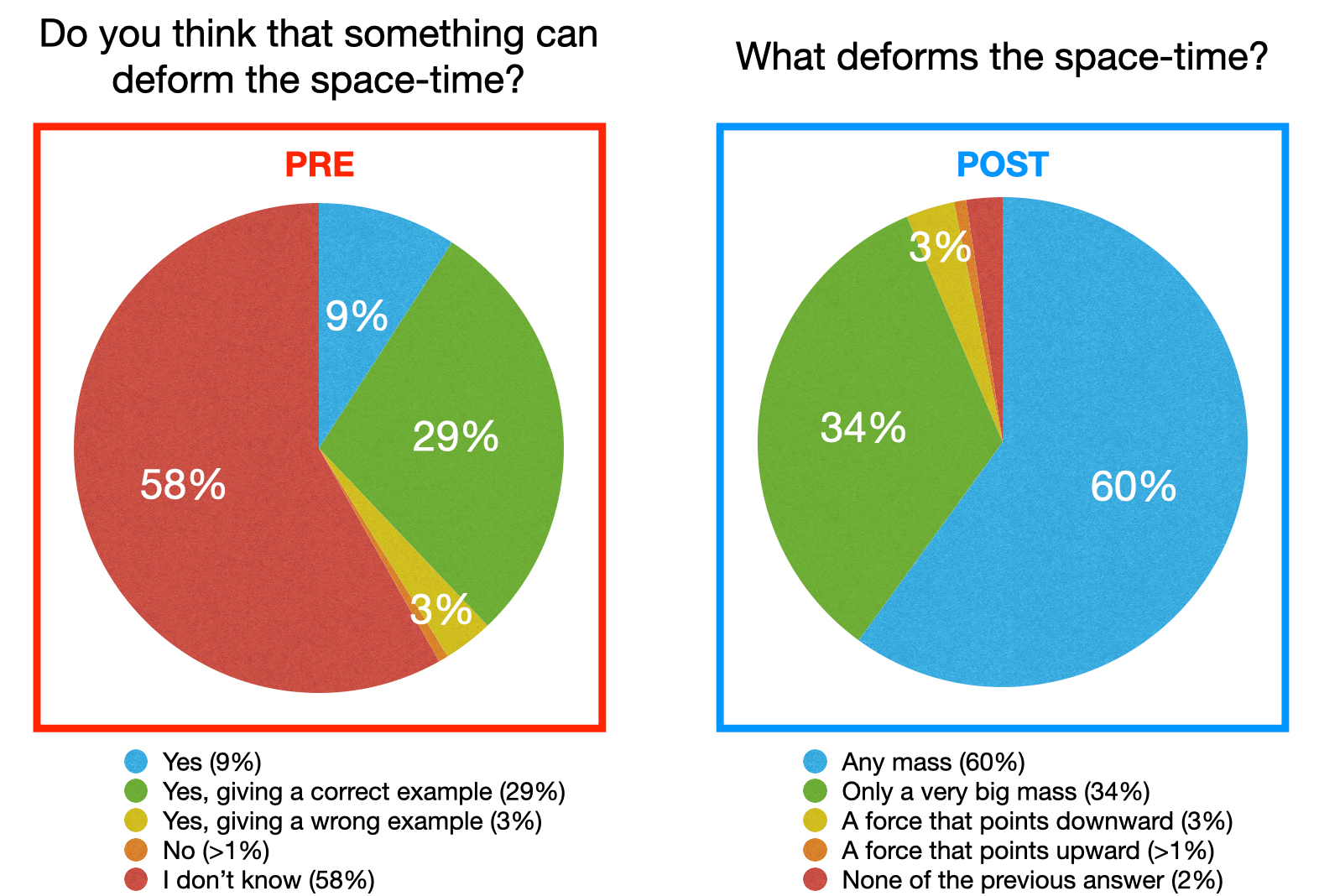}
\caption{The distribution of the answers to the questions related to the cause of the deformation of the space-time. The percentage of each answer is shown in the legend in brackets (left: pre-questionnaire, right: post-questionnaire).}
\label{fig:q3}
\end{figure}

\begin{figure}[ht]
  \centering
\includegraphics[width=0.9\columnwidth]{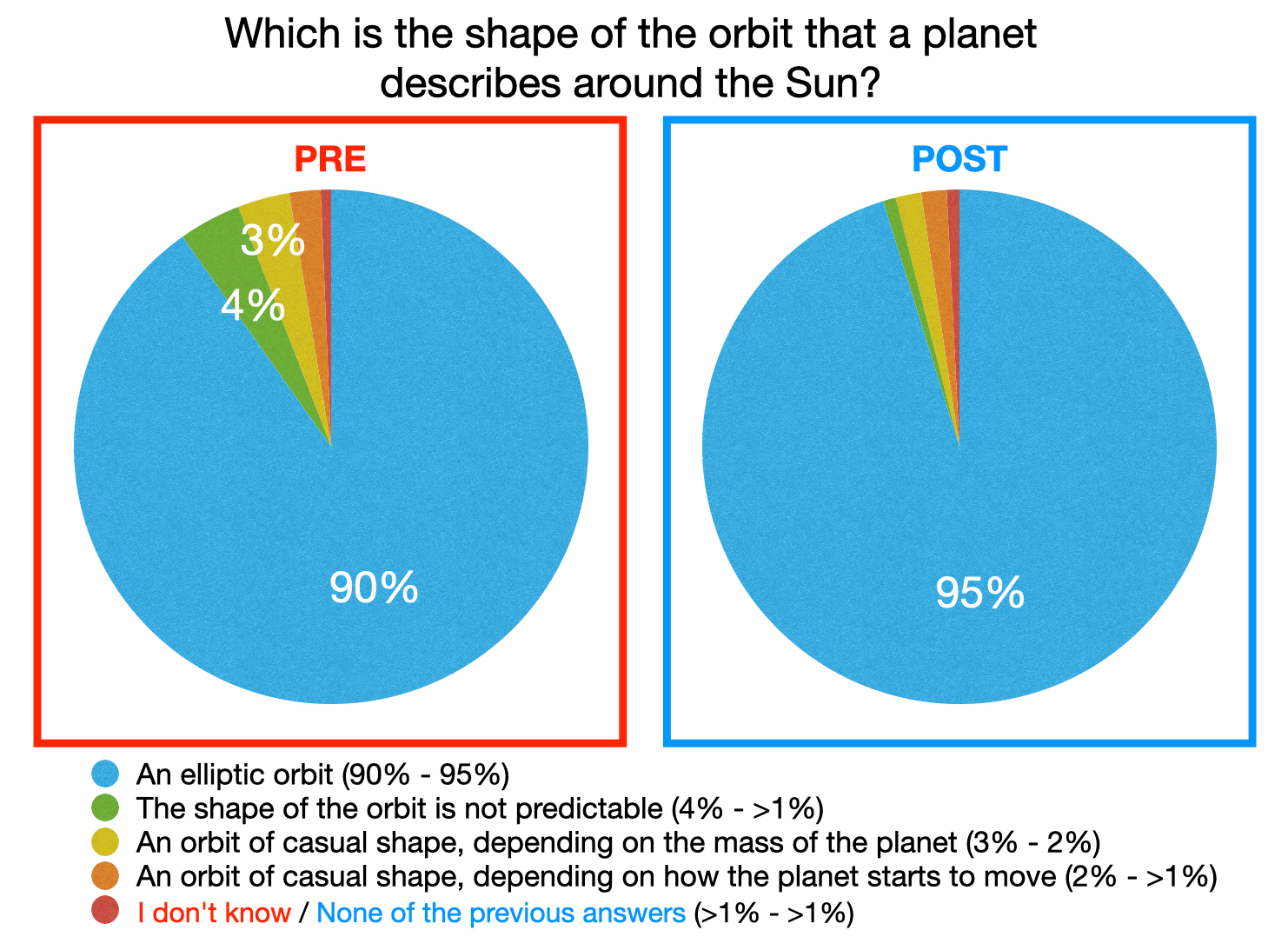}
\caption{The distribution of the answers to the question \textit{Which is the shape of the orbit that a planet describes around the Sun?}. The percentage of each answer is shown in the legend in brackets (left: pre-questionnaire, right: post-questionnaire).}
\label{fig:q4}
\end{figure}

\begin{figure}[ht]
  \centering
\includegraphics[width=0.98\columnwidth]{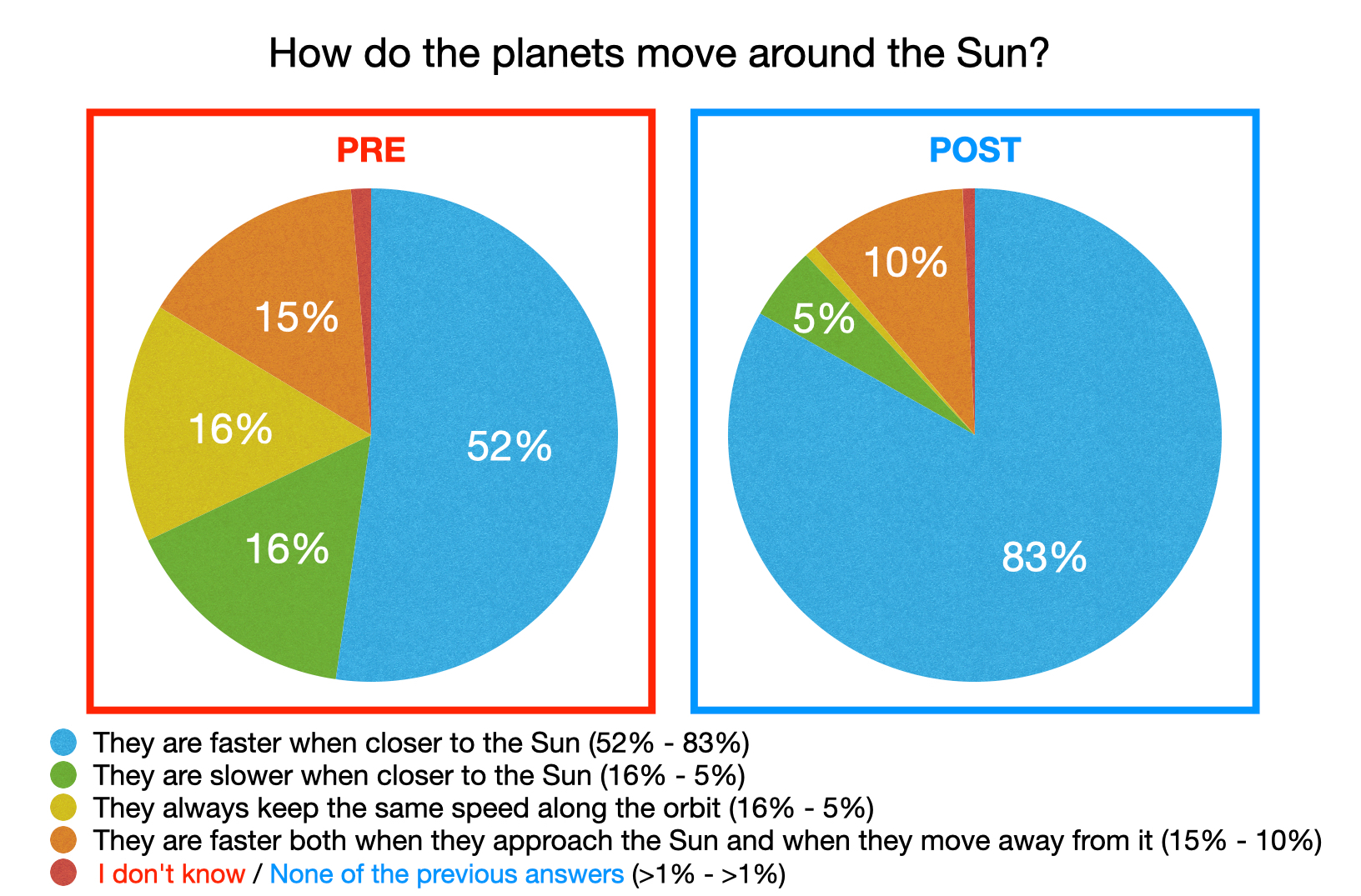}
\caption{The distribution of the answers to the question \textit{How do the planets move around the Sun?}. The percentage of each answer is shown in the legend in brackets (left: pre-questionnaire, right: post-questionnaire).}
\label{fig:q5}
\end{figure}

\begin{figure}[ht]
  \centering
\includegraphics[width=0.93\columnwidth]{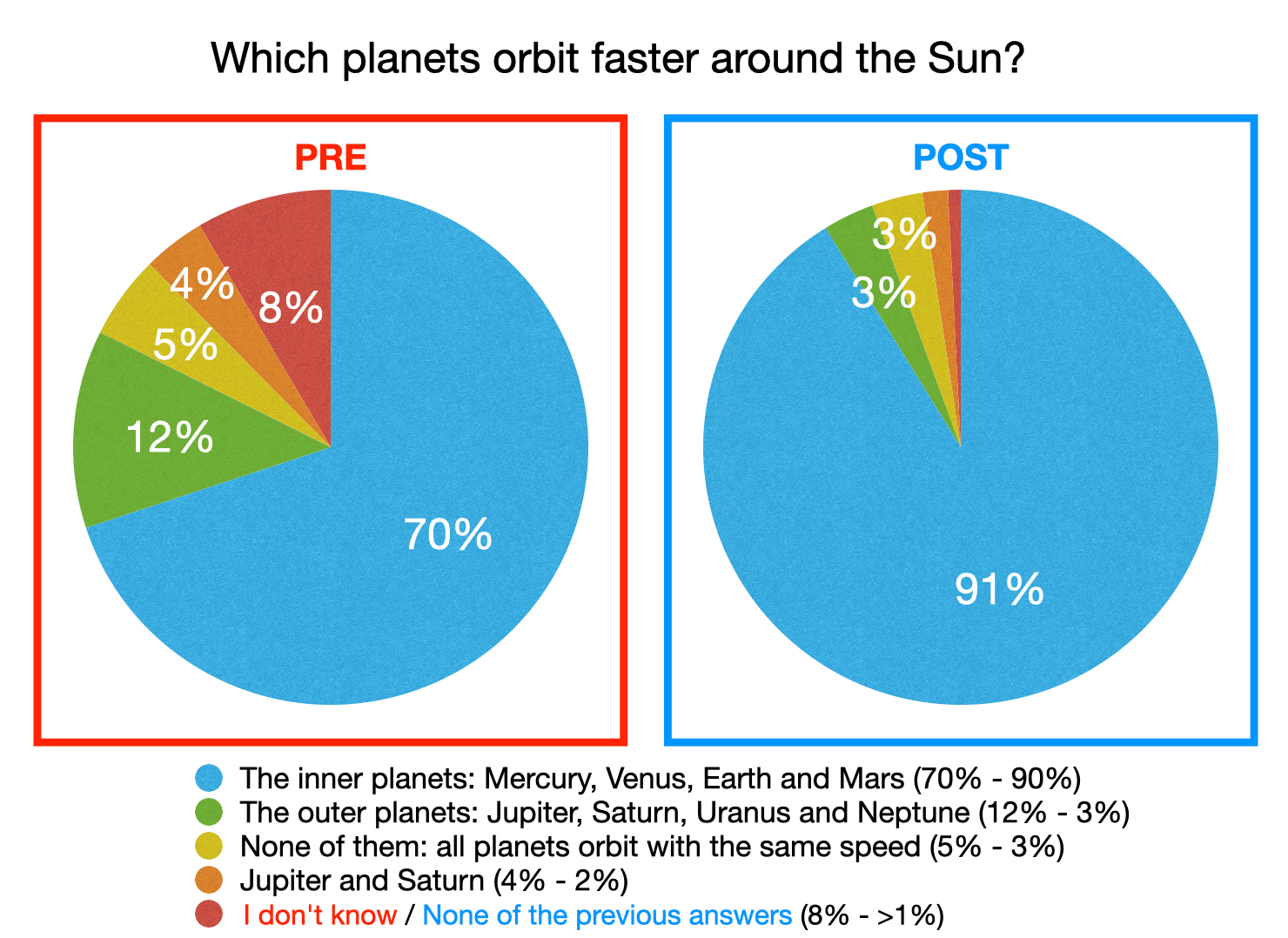}
\caption{The distribution of the answers to the question \textit{Which planets orbit faster around the Sun?}. The percentage of each answer is shown in the legend in brackets (left: pre-questionnaire, right: post-questionnaire).}
\label{fig:q6}
\end{figure}

\begin{figure}[ht]
  \centering
\includegraphics[width=1\columnwidth]{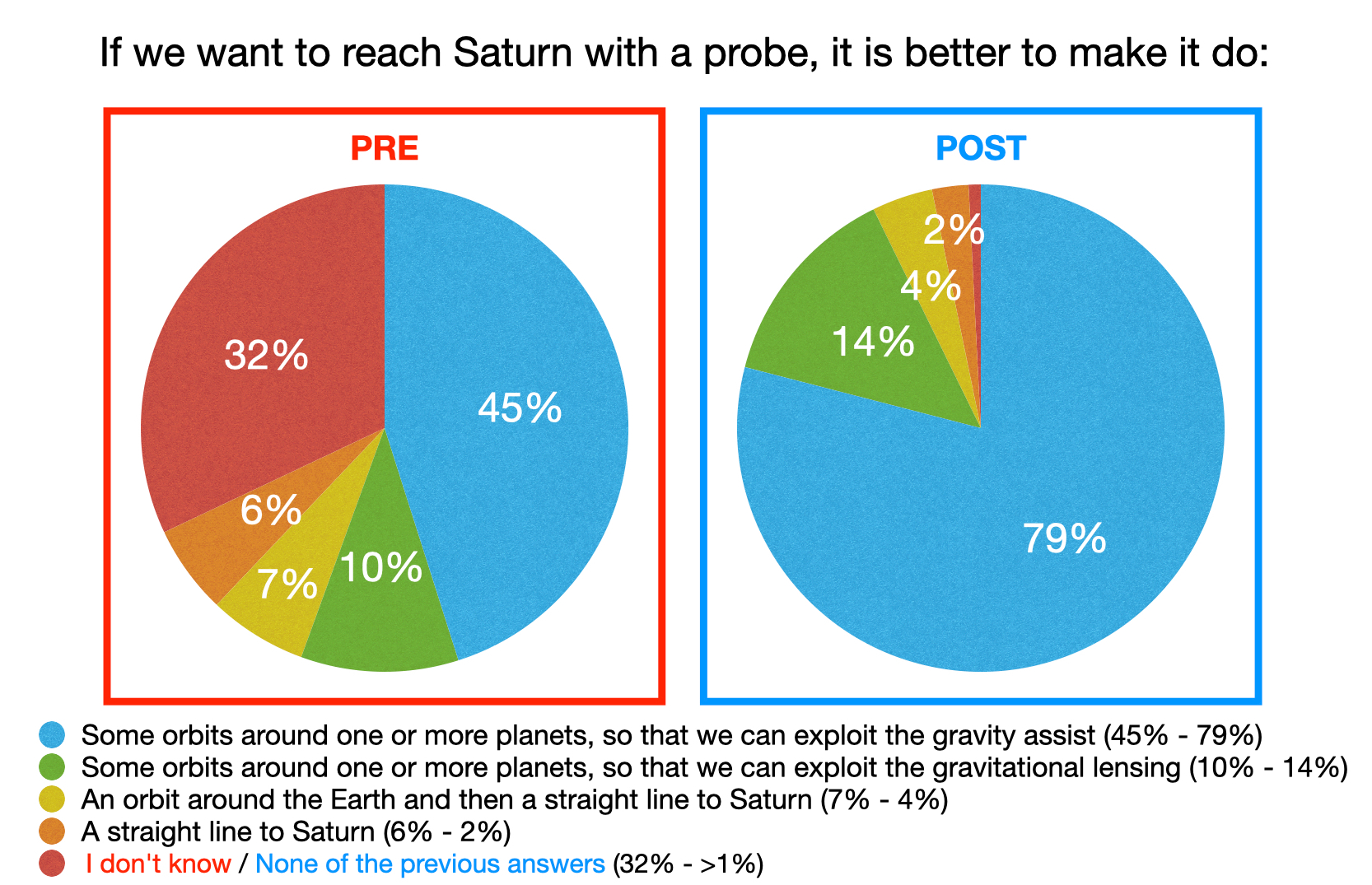}
\caption{The distribution of the answers to the question \textit{If we want to reach Saturn with a probe, it is better to make it do}. The percentage of each answer is shown in the legend in brackets (left: pre-questionnaire, right: post-questionnaire).}
\label{fig:q7}
\end{figure}

\begin{figure}[ht]
  \centering
\includegraphics[width=1\columnwidth]{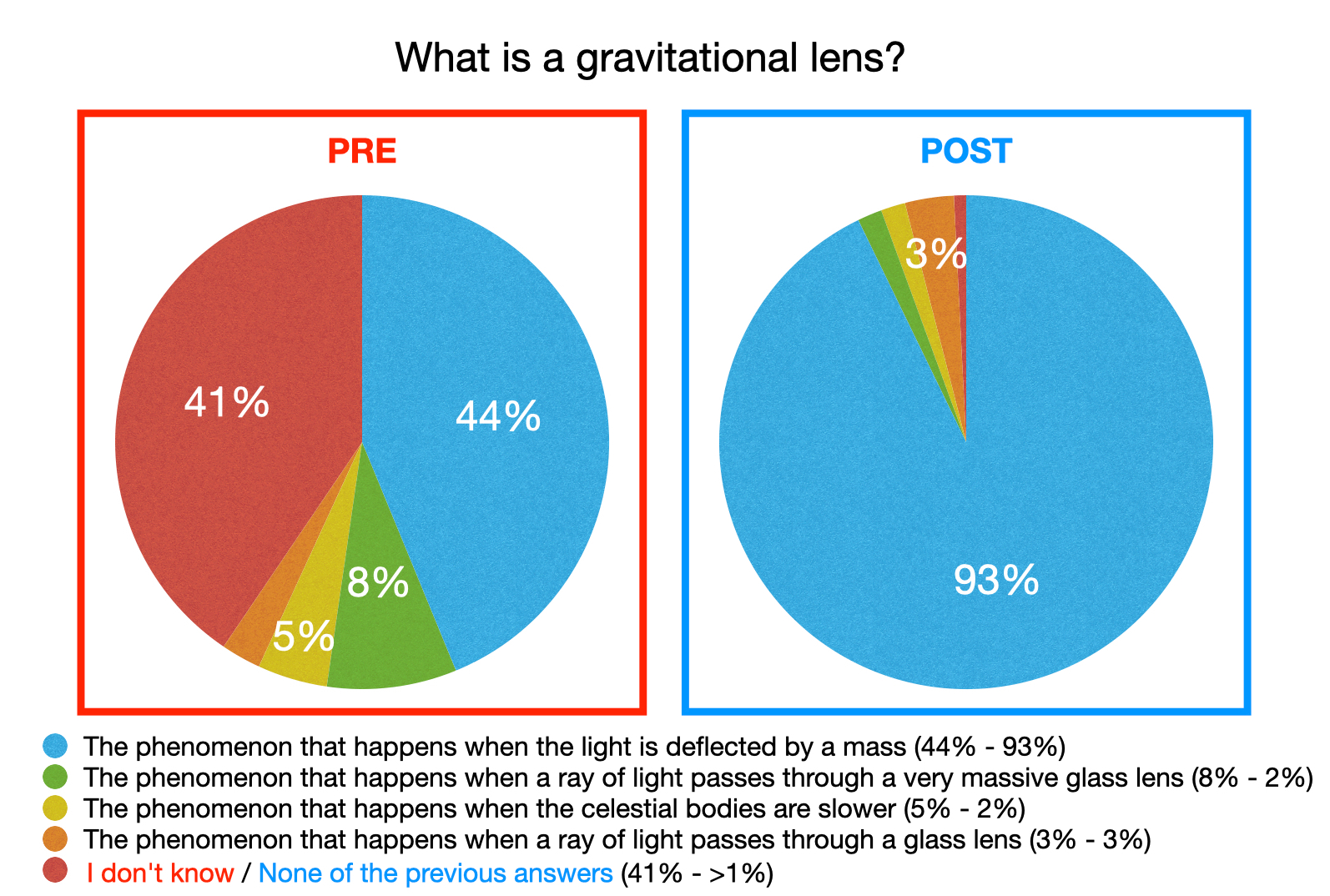}
\caption{The distribution of the answers to the question \textit{What is a gravitational lens?}. The percentage of each answer is shown in the legend in brackets (left: pre-questionnaire, right: post-questionnaire).}
\label{fig:q8}
\end{figure}

\begin{figure}[ht]
  \centering
\includegraphics[width=1\columnwidth]{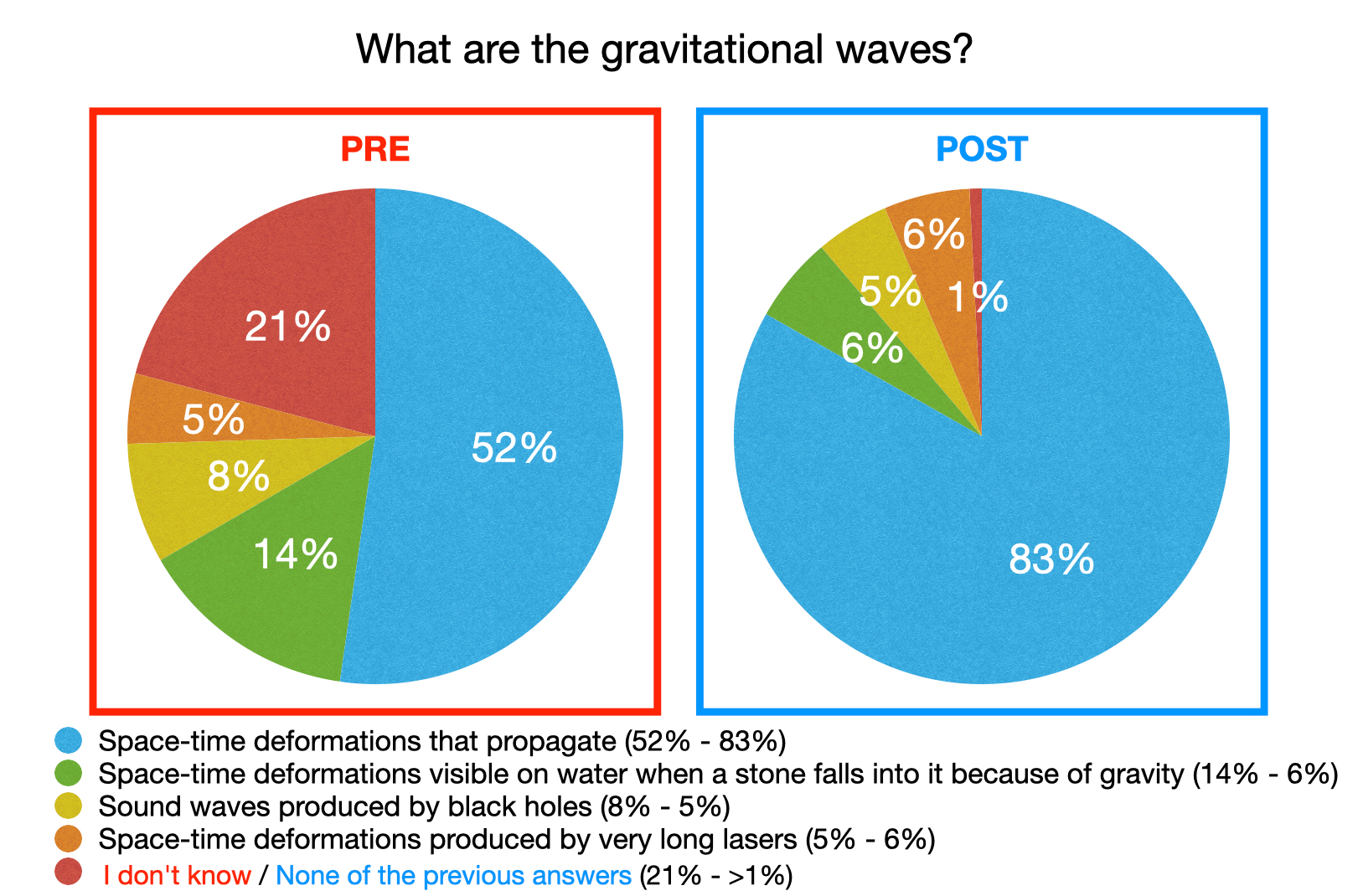}
\caption{The distribution of the answers to the question \textit{What are gravitational waves?}. The percentage of each answer is shown in the legend in brackets (left: pre-questionnaire, right: post-questionnaire).}
\label{fig:q9}
\end{figure}

\begin{figure}[ht]
  \centering
\includegraphics[width=1\columnwidth]{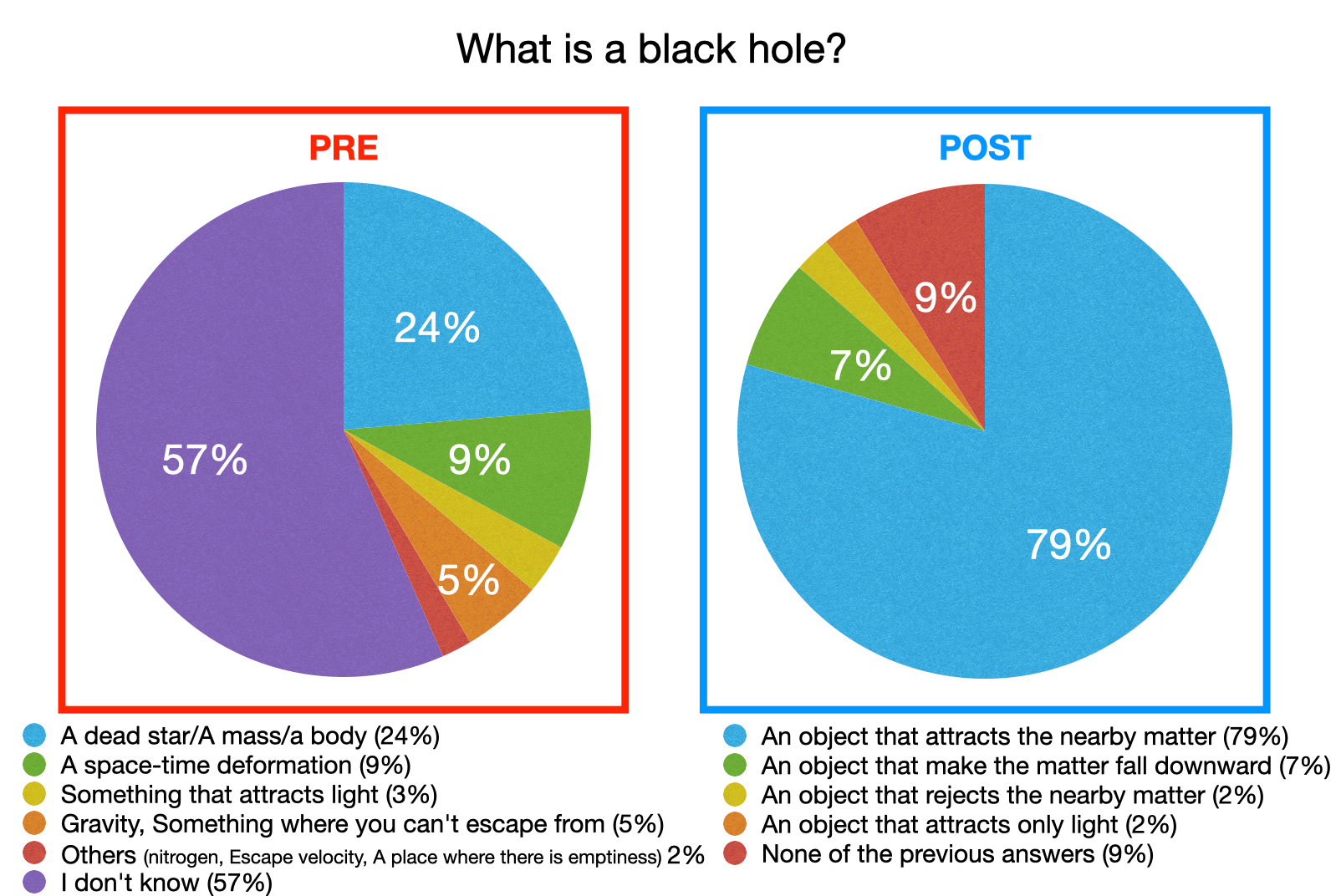}
\caption{The distribution of the answers to the question \textit{What is a black hole?}. The percentage of each answer is shown in the legend in brackets (left: pre-questionnaire, right: post-questionnaire).}
\label{fig:q10}
\end{figure}

\begin{figure}[ht]
  \centering
\includegraphics[width=1\columnwidth]{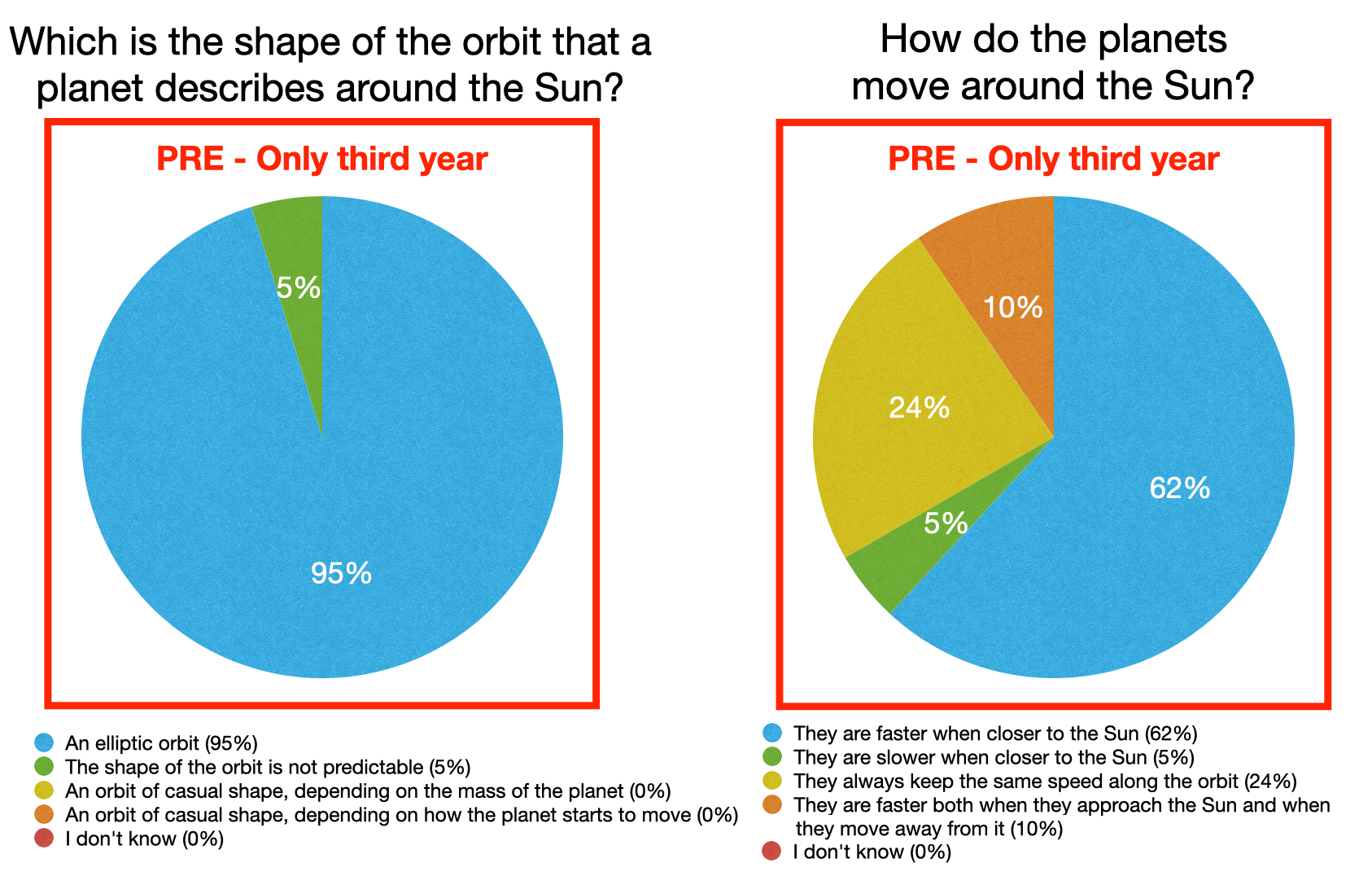}
\caption{The distribution of the answers to the questions related to Kepler's second and third law by participants attending the third year of high school, before the activity. The percentage of each answer is shown in the legend in brackets.}
\label{fig:terzo_anno_pre}
\end{figure}

\begin{figure}[ht]
  \centering
\includegraphics[width=1\columnwidth]{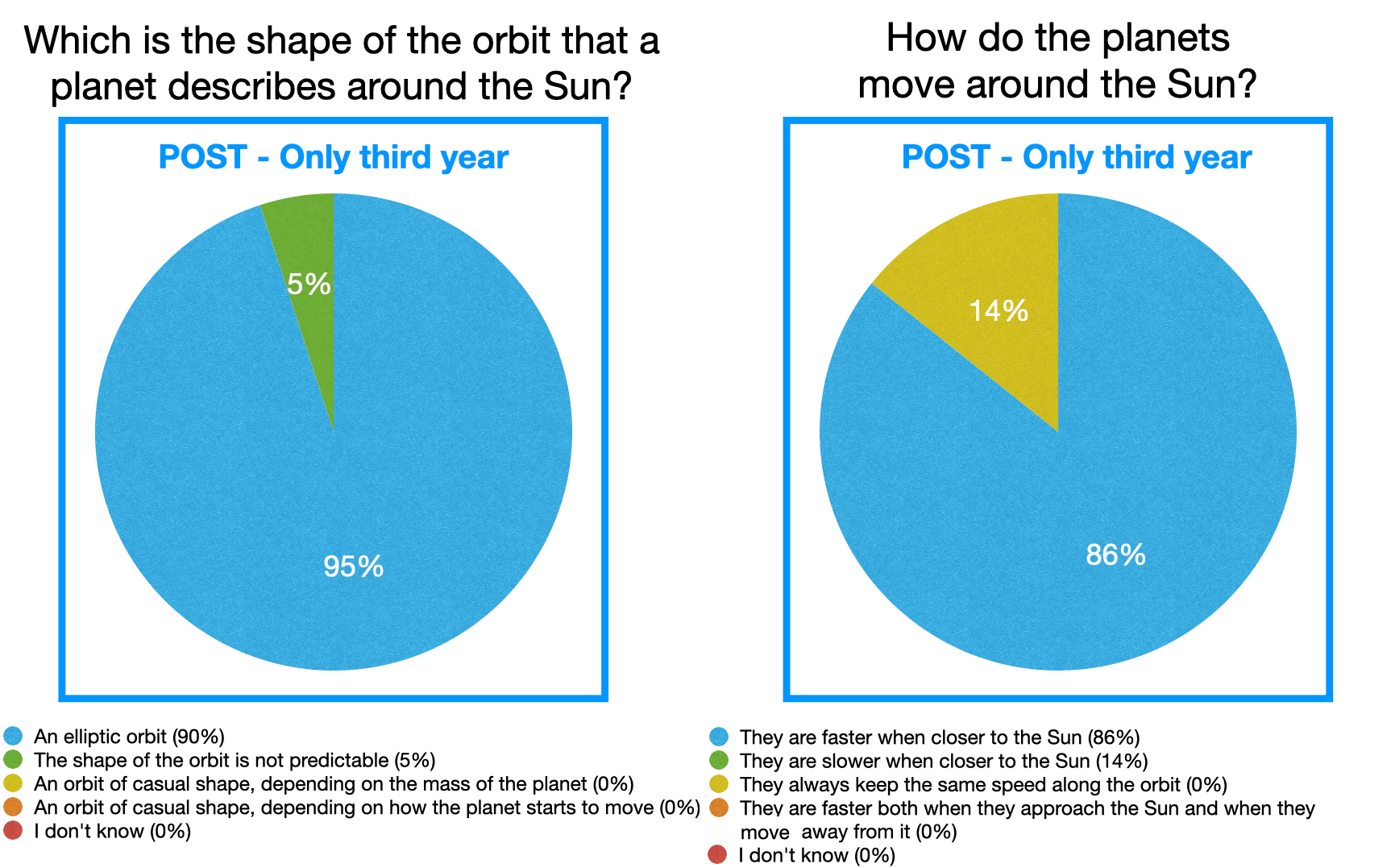}
\caption{The distribution of the answers to the questions related to Kepler's second and third law by participants attending the third year of high school, after the activity. The percentage of each answer is shown in the legend in brackets.}
\label{fig:terzo_anno_post}
\end{figure}

\clearpage
\newpage
Given this large difference of the sample between the first two and the third questionnaire, this latter will be taken into consideration only as a general trend and thus the corresponding data can be found in the Appendix. 

All the participants were attending the scientific high school (\lq\lq Liceo Scientifico\rq\rq), where Newtonian gravity is typically studied starting from the third year. The age of the participants ranges from the first year of the Italian high school (13-14 years old) to the fifth and last year. As shown in \textbf{Figure \ref{fig:q0}}, the majority of the participants already treated the Newtonian gravity in school. 

\subsection{What is the space-time and how it is deformed}
To assess the participants' understanding of the concept of space-time and its deformation, we asked for a definition of space-time (\textbf{Figure \ref{fig:q1}}). Before the activity, the 46\% of the students' answers is correct (\textit{It's the union of space and time, who together form a 4-dimensional entity}), while the 34\% of them states that it is 2-dimensional. The remaining 20\% of the participants confuses the space-time or with the fourth dimension, or with the speed (which in school is usually defined as the ratio between space and time) or says they do not know. After the activity, the percentage of correct answers increases: over 90\% of the participants states that the space-time has 4 dimensions, while only the 6\% still confuses the concept of space-time with the fourth dimension. After four months this positive trend is confirmed, since the percentage of correct answers is 86\% (see the Appendix).
 
We then focused on the cause of the deformation (\textbf{Figure \ref{fig:q3}}). Before the activity with the rubber sheet, the majority of respondents (58\%) says that they do not know if there is something that can deform the space-time, while the 1\% states that nothing can deform it, while the 41\% declares that it can be deformed. In more detail, the 9\% generically states that it can be deformed, the 28\% gives a correct example of the source of the deformation, while the 4\% gives a wrong example. 

Immediately after the activity, the 60\% of the participants states that the cause of the deformation can be any mass, the 34\% only refers to a big mass while the 3\% cites a force that points downward. The same trend can be found four months after the activity, when the answer \lq\lq big mass\rq\rq  is cited by the 31\% of the participants (see the Appendix). 

\subsection{Kepler's laws and gravity assist}
Later, we investigate the participants' knowledge of the three Kepler's laws.

Before the activity, the majority of the students already remember the first Kepler’s law, since the 90\% states that the shape of the orbit that a planet describes around the Sun is elliptical (\textbf{Figure \ref{fig:q4}}). The 4\% says that planets have a random orbit, while only less than 1\% admits they do not know. Immediately after the activity the situation slightly improves, since the percentage of correct answers stays high (95\%), while the remaining 5\% of the participants gives wrong answers. Even after four months, the 93\% of the sample gives the correct answer (see the Appendix).

The Kepler's second law shows a more considerable improvement (\textbf{Figure \ref{fig:q5}}). Before the activity, in fact, just over half (52\%) of the participants correctly answer when asked about the way in which planets move around the Sun. The others (48\%) think that the planets move slower when they are closer to the Sun (16\%), that they always keep the same speed along the orbit (16\%) and that they are faster both when they approach the Sun and when they move away from it (15\%). Right after the activity the percentage of correct answers rises to 83\%, while all the other wrong answers reduce. After four months from the activity about the 80\% of the sample of respondents still remember the correct answer (cfr. Appendix). 

Regarding Kepler's third law, participants were asked which planets orbit faster around the Sun (\textbf{Figure \ref{fig:q6}}). If before the activity the 70\% already know how to correctly respond, this percentage rises to 91\% after the activity, and to 83\% four months after (see the Appendix). 

In order to investigate the comprehension of the gravity assist, we asked how space probes travel in space (\textbf{Figure \ref{fig:q7}}). In this case the 55\% of the participants do not correctly answer the related question of the pre-questionnaire. After the activity the majority of students (79\%) have understood what is gravity assist, a trend that is confirmed after four months (76\%).  
\subsection{Gravitational lensing, waves and black holes}
Before the activity, only less than a half of the participants (44\%) gives the right answer when asked to describe the phenomenon of gravitational lensing (\textbf{Figure \ref{fig:q8}}), while the 41\% declares they do not know at all. The remaining 16\% confuses the phenomenon or with something related to a glass lens, or explain the phenomenon using the Italian word \lq\lq lens=lenti\rq\rq meaning \lq\lq slow\rq\rq instead of lens. 
After the activity, the correct answers rise to 93\% of the total, while the percentage of those who associate the phenomenon to a glass lens remains unchanged (3\%). 
Also after four months the percentage of correct answers confirms the improvement of the result (98\% of correct answers).

Regarding the gravitational waves, before the activity, about 52\% gives the correct definition of them (\textbf{Figure \ref{fig:q9}}). The remaining half is made up of a 21\% who admits they do not know what they are, and a 27\% who gives wrong answers confusing them with other topics they studied in school (waves formed on water) or with concepts they have heard of after the spread of the news about discovery (black holes and lasers). After the activity the 83\% gives the correct definition. After four months we have a percentage of 81\% of correct answers. 

Finally, we asked the participants to give us a definition of black holes (\textbf{Figure \ref{fig:q10}}). Before the activity, the 57\% does not feel confident enough to give an answer. Only less than the 2\% gives completely wrong answers (the escape speed, nitrogen, a place where there is an emptiness). The remaining 41\% gives an answer not far from the correct one: the 24\% associates black holes with a mass, a body or a dead star, the 9\% with a deformation of space-time, the 3\% with something that absorbs everything, or nearby matter, or energy or light, the 5\% associates them with a very intense gravity or with something that cannot be escaped. After the activity, the percentage of correct answers rises to 79\%, giving that they associate black hole with something that attracts only the nearby matter. Only the 7\% chooses the answer inserted as a distractor, i.e. that the black hole attracts matter \textit{downward}. This is also confirmed by the trend shown in the four months later answers: the 84\% gives the right answer, while only the 7\% recalls the idea of an object that attracts \textit{downward}. 

\section{Discussion}
Overall, the students who participated in our activity with the rubber sheet have significantly improved their knowledge about the topics addressed. This trend is also confirmed by the responses obtained four months after the activity. 

As regards the topics usually treated in school, the activity straightens their knowledge and improves their understanding. The most significant case in this sense is Kepler's second law, that the rubber sheet has helped to visualize and remember. In particular, while the first law was already known before the activity by the great majority of students (90\%), the second was remembered only by the 52\% of the total. This trend could suggest that the second law is studied in school in a more mnemonic way with respect to the first law, without sufficient understanding. In fact, if we consider only the students attending the third year (who have just treated the Kepler’s laws in school, and thus who should remember them better), we note that only the 62\% of them correctly recall the second law before the activity, while the first law reaches a percentage of the 95\% (\textbf{Figure \ref{fig:terzo_anno_pre}}). 

Moreover, when verbally asked during the activity, the participants often remembered the statement of the second law \lq\lq \textit{A line segment joining a planet and the Sun sweeps out equal areas during equal intervals of time.}\rq\rq \ but they evidently did not understand its meaning. Only when they saw the motion of the marbles on the sheet they really understood it, as confirmed by the results of the post-questionnaires. 
In fact, after the activity, also the percentage of correct answers for the second law becomes very high, as shown in \textbf{Figure \ref{fig:terzo_anno_post}} (86\%). 

Regarding the topics that are not typically treated in school, like black holes, gravitational waves and gravitational lense, the data show a prior knowledge lower than the one related to Kepler’s laws, as might be expected.

Despite this fact, also for these topics the improvement has been remarkable, since the percentage of correct answers increases from less then a half of the total to almost the 100\%. Our analysis also shows that the acquired knowledge seems to be lasting, given that almost all of the respondents of our sub-sample continue to correctly answer even after four months from the activity. Moreover, it can be seen that, although these topics are not addressed in high school curricula, they were not completely unknown by the students even before the activity. This could be a sign of a widespread fascination for these topics, which encourage the students to search for information even if they are not treated in school (as indicated by the fact that they give reasonable answers even before the activity: black holes are \lq\lq dead stars\rq\rq, \lq\lq space-time deformations\rq\rq, \lq\lq something where you can’t escape from\rq\rq), or at least of a strong bond with current news (as suggested by the answers we received for the gravitational waves, related to lasers or black holes).

Our analysis also suggests that the rubber sheet can also be effective in dealing with the misconceptions and wrong beliefs related to gravity. In particular, our data allow us to investigate two misconceptions: the fact that the deformation of space-time is only due to big masses and the idea that gravity is always a force that points downward.
As regards the former, the data shows that our activity, even having faced this aspect, does not solve it completely, since there is a substantial part of students (34\%) that after the activity refers only to very big masses when asked about the source of deformation. This means that a greater attention must be paid in fighting this idea throughout the activity with the rubber sheet, emphasizing more than once that space-time is not only related to stars and galaxies, but also to objects with smaller mass, such those who populate our everyday life.

Regarding the idea of a gravity that points downward, our data show that it can indeed be fought using the RSA, since only the 3\% of our sample choose the distractor \lq\lq a force that points downward\rq\rq \  when asked to specify what deforms the space-time. This means that the discussion that is made at the beginning of the activity, that deals with the absence of a privileged direction of gravity outside the Earth, and with the limitations of the rubber sheet analogy, has paid off. Our idea is also confirmed by the results obtained after four months from the activity, when the answer “downward” completely disappears, and also by the results obtained for what concerns black holes.

Overall, our analysis therefore leads us to think that the rubber sheet, albeit with all its limitations, represents a formidable tool for introducing GR at high school level, given that it helps students both to visualize how gravity works and to remember it longer.  

\section{Summary and conclusion}
In this paper we presented the results of the questionnaires we have administered to over 150 high school students who participated in the activity held at the Department of Mathematics and Physics of Roma Tre University that uses the rubber sheet analogy to address several topics related to gravity.
Our data show that the rubber sheet can indeed be very useful not only in treating topics that can be explained by GR, but also to better understand and remember the topics generally addressed in high school using Newtonian theory of gravity.  

In the next future we plan to continue to test our proposal involving an increasing number of students. In particular, we hope to resume the activity in person, so that we could also follow more closely the compiling of the questionnaire administered four months after the activity, in order to have statistics comparable to the ones obtained with the questionnaires distributed right before and right after the activity. This will help us to understand and quantify better the long-term effect of introducing the rubber sheet at high school level.

\section*{Acknowledgements}
This work has been supported by the Italian Project 'Piano Lauree Scientifiche' and the Young Minds Section of Rome of the European Physical Society. A special thanks goes to the mechanical shop of the INFN Roma Tre Section that built our aluminum space-time structure. Also thanks to the teachers and students who participated in our activity at the Department of Mathematics and Physics of Roma Tre University with their classes.

\newpage
\section*{Appendix}

\begin{figure}[ht]
  \centering
\includegraphics[width=0.83\columnwidth]{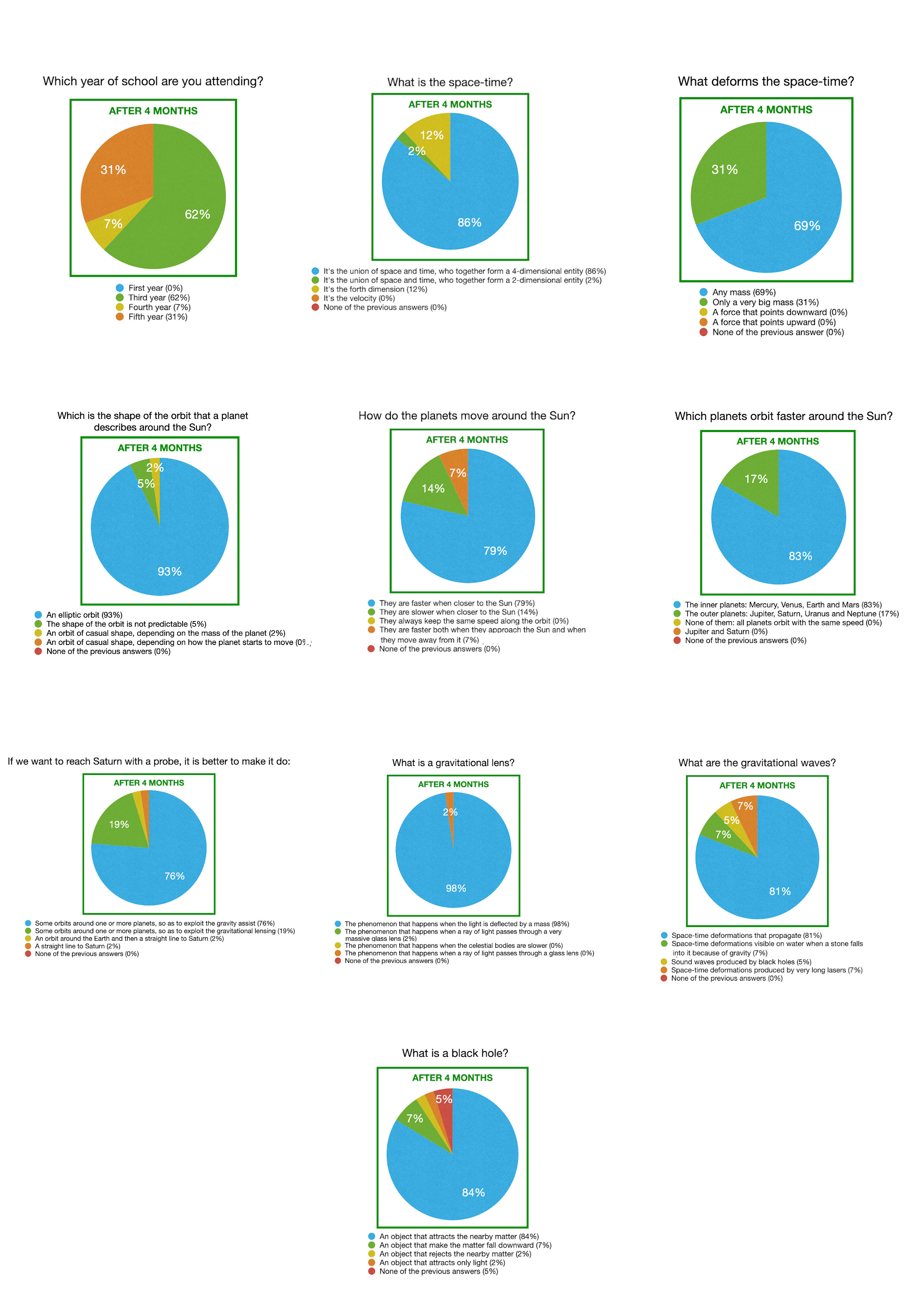}
\caption{The results to the questionnaire administrated four months later the activity.}
\label{fig:postpost}
\end{figure}

\end{document}